%% file: astroph_results_7_2.tex
\documentclass[aps,preprint,floats,epsfig,tightenlines]{revtex4}

\usepackage{graphicx}

\def\simgt{\ \raisebox{-.25ex}{$\stackrel{>}{\scriptstyle \sim}$}\ }

\begin{document}

\title{Limits on the Ultra-High Energy Electron Neutrino Flux from 
the RICE Experiment\\
\vskip .5cm
}
\date{\today}
\input riceauthor
\begin{abstract}
Upper limits are presented on the diffuse flux of ultra-high energy 
$\nu_e$, based on analysis of data taken by the RICE experiment during 
August, 2000. The RICE receiver array at South Pole monitors cold ice for 
radio-wavelength Cherenkov radiation resulting from neutrino-induced 
in-ice showers. For energies above 1 EeV, RICE is an effective detector 
of over 15 km$^3$ sr. Potential signal events are separated from 
backgrounds using vertex location, event reconstruction, and signal 
shape. These are the first terrestrial limits exploiting the physics of 
radio Cherenkov emissions from charged-current $\nu_e+N\to e+N'$ 
interactions. 
\end{abstract}
\maketitle


\section{Introduction}
\label{intro.sec}

Neutrinos are generally expected to be an important component of 
ultra-high energy (UHE) cosmic rays\cite{GaisserHS}. Neutrinos propagate 
in straight lines from their source and retain spectral information about 
their origin - in contrast to protons, electrons or gamma-rays, which 
interact with cosmic magnetic fields, cosmic radiation backgrounds at 
microwave and radio frequencies, and material within the source itself. 
Energies below 100 GeV are the province of solar, supernova and 
atmospheric neutrino experiments. Above 100 GeV it is expected that 
astrophysical sources (e.g. AGN, GRBs, the galactic disk) will be 
observable in planned cubic kilometer scale 
detectors\cite{IceCube,ANTARES,NEMO,NESTOR}, or perhaps in prototype 
existing experiments that have pioneered the development of large arrays 
of optical Cherenkov detectors\cite{AMANDA,BAIKAL}. Despite their 
impressive size, even these experiments may be too small to see 
scientifically interesting fluxes at energies above 10 PeV. Of particular 
interest, neutrinos of energy 0.1-1 EeV may provide key information for 
understanding anomalies reported in the cosmic ray 
spectrum\cite{AGASA,HPARK,FLYSEYE} above the ``GZK" cutoff\cite{GZK}. 
Detection of the nominal flux of GZK neutrinos\cite{BerZat69} requires a 
detector mass equivalent to 100 km$^3$ of water\cite{EngelSS}. To achieve 
such large masses, other techniques have been proposed for this energy 
range, including detection of horizontal air 
showers\cite{AUGER,CapelleCPZ,OWL,EUSO}, acoustic signals in water or 
ice\cite{acoustic} and a variety of ideas\cite{MarkovZ, Zheleznykh, 
radhep} that rely on detection of radio Cherenkov emission via a process 
first proposed by Askaryan\cite{Askaryan} and recently confirmed in the 
laboratory\cite{Saltzberg}. The latter set includes detection 
of radio pulses generated in the lunar 
regolith\cite{GorhamLN,GorhamLNSW}, geological salt deposits\cite{wipp}, 
and Antarctic ice as observed {\it in situ}\cite{FrichterRM} or from a 
high altitude balloon\cite{ANITA}. We report here on the first results 
from the Radio Ice Cherenkov Experiment (RICE), an array of dipole 
antennas located in-ice at the South Pole. 

\begin{figure}[htpb]
\centerline{\includegraphics[width=10cm]{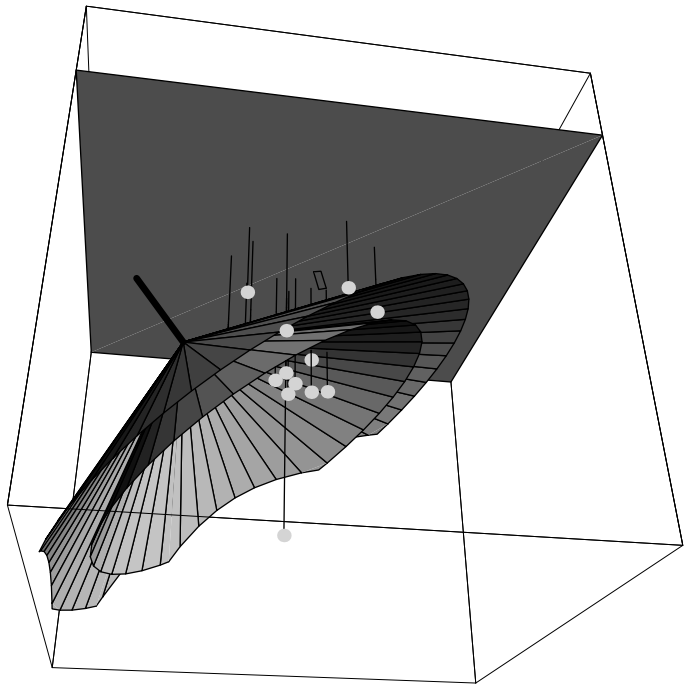}}
\caption{
RICE concept.  The RICE array (viewed from below) is drawn schematically 
to scale, showing cables and antenna locations. The small rectangle on 
the surface is the Martin A. Pomerantz Observatory (MAPO) which houses 
the RICE electronics. An electron neutrino is incident from above the 
horizon and interacts in the ice. An electromagnetic shower (hidden) 
results, which in turn generates an outgoing cone of radio Cherenkov 
radiation. Two cones, drawn at $\theta_c \pm 3^\circ$, depict an 
approximate 3 dB range of signal strength.
}
\label{fig:concept}
\end{figure}

The concept behind RICE is illustrated in Figure \ref{fig:concept}: a 
neutrino-nucleon interaction results in an electromagnetic cascade, which 
in turn produces a few nanosecond radio pulse with power concentrated 
around the Cherenkov angle. Detection of that pulse by several receivers 
with fast electronics allows for reconstruction of the interaction vertex 
and the energy/direction of the incident neutrino. Development and status 
of the RICE experimental configuration is summarized in Section 
\ref{sec:overview}.

The experiment requires accurate simulation of the expected signal, full 
calibration of an instrument with many interacting components, and 
development of techniques for event reconstruction and elimination of 
spurious backgrounds. Simulation and calibration are discussed in Section 
\ref{sec:sigpath}. Additional information concerning our use of numerical 
techniques to model the Cherenkov signal may be found in 
Refs.~\cite{RazzaqueSBMRS,RazzaqueSBMRSa}. Calibration of the RICE 
experiment is described in greater detail in Ref.~\cite{Kravchenko}.

Event reconstruction (Section \ref{sec:recon}) is important not only for 
the eventual analysis of neutrino observations, but also as a tool for 
separation of neutrino events from backgrounds. Background events result 
from either correlated or uncorrelated receiver hits. The former are due 
to impulsive transients associated with human activity in and around 
South Pole station\cite{Frichter99}. Uncorrelated receiver hits result 
from thermal noise in the ice. The presence of continuous wave 
transmissions is readily detected, and is accounted for in the data 
analysis. Related issues are discussed in Section \ref{sec:back}.

In development since 1995, the bulk of RICE operations has been devoted 
to calibration, understanding backgrounds, and developing strategies for 
dealing with those backgrounds. As a result, the experimental 
configuration has been dynamic - a situation which complicates both 
data-taking and analysis. We therefore chose to ``freeze" the 
configuration for one month (August 2000), take data, and analyze the 
results. The analysis of August 2000 data is presented in 
Section~\ref{sec:augytwok}. Refinement of the techniques used in the 
present analysis has been an ongoing process. As such, the present 
analysis is more complete than the preliminary analysis presented in 
Ref.~\cite{Seckel01} and our present limit is somewhat stronger, although 
based on the same data.

Limited to a single month of data, the exposure analysed here is too 
short to provide a significant likelihood of finding actual neutrino 
events, at least given theoretical predictions. Our current search yields 
a null result, and we use it to place upper limits on proposed 
theoretical models. Accordingly, we have taken a conservative approach 
throughout our analysis, generally choosing simpler options when 
confronted with choices in the analysis, often resulting in a less 
restrictive limit. We present a broad survey of systematic uncertainties 
in Section~\ref{sec:disc}.

RICE was originally conceived as an experiment to search for electron 
neutrinos via charged current interactions and the subsequent 
electromagnetic showers. Since these are reasonably well studied, the 
signal predictions are fairly robust and these events remain the primary 
subject of the current paper. It is understood, however, that at energies 
well in excess of 1 PeV, LPM effects reduce the efficiency for detecting 
electromagnetic showers. As a result, hadronic recoil showers, which are 
produced by neutrinos of all flavors in both charged and neutral 
currents, may be important for detection of neutrinos at the highest 
energies. Section~\ref{sec:disc} includes a discussion of 
the efficiency for RICE to detect neutrinos via hadronic showers.

Finally, we summarize our discussion in Section~\ref{sec:summary} and 
consider possible future applications of the radio technique at the South 
Pole.


\section{Overview of RICE}
\label{sec:overview}

During August 2000, the RICE experiment consisted of a 16-channel array 
of dipole radio receivers (``Rx''), deployed within a 200 m$\times$200 
m$\times$200 m cube, at 100-300 m depths. The 10-cm fat dipole antennas 
were designed both to produce a peak response at 250 MHz (compatible with 
our expected signal and the rest of the DAQ), and to allow piggyback 
deployment in holes drilled for the AMANDA experiment. Signals from each 
antenna are boosted by an in-ice amplifier (36 dB), carried by coaxial 
cable to the surface observatory, filtered to suppress noise below 200 
MHz, re-amplified (52 or 60 dB), and fed into a CAMAC crate. The gain of 
the in-ice amplifier is chosen so as to allow measurement of the in-ice 
radio background ($T_N \sim 250$K), after transmission over the co-axial 
cable. The gain of the surface amplifier is chosen to bring the in-ice 
thermal background and expected signals within the dynamic range of the 
DAQ electronics. The amplifiers are adjusted for each channel so as to 
reduce the channel-to-channel variation in receiver response. The 200 MHz 
filter effectively eliminates low frequency incoherent galactic radio 
noise, and an impulsive background generated by AMANDA phototubes which 
peaks below 50 MHz. The RICE experimental bandpass is effectively $\Delta 
f \approx 200-500$~MHz.

After entering the CAMAC crate, the signals are split, one copy being 
sent to a bank of 4 digital oscilloscopes, and the other going to trigger 
logic. The trigger requires a four-fold coincidence within a 1.2 $\mu$s 
window, four hits being required to perform event reconstruction. A 
common discriminator threshold is adjusted to maintain a trigger rate 
that permits monitoring of backgrounds without a significant reduction in 
livetime for true neutrino events. For most event geometries, the 1.2 
$\mu$s window accomodates signals propagating across the array and along 
cables of differing length to reach the surface DAQ.

In the event of a trigger, times for each channel to cross the 
discriminator threshold are recorded by a TDC module in the DAQ. These 
times form the basis for an on-line software veto designed to reject 
noise events resulting from surface activity.  Events passing the veto 
are recorded with 8$\mu$s of waveform data per channel transferred from 
the scope and saved on disk for later analysis. The waveform data is 
critical to understanding the background environment, the nature of noise 
events, and provides the event timing used for performing event 
reconstruction.

In addition to the receivers, four antennas are configured and deployed 
as in-ice transmitters for use as calibration sources. Three horn 
antennas are deployed on the surface, and provide a further veto of 
surface activity. More details of the DAQ and trigger configuration can 
be found in \cite{Kravchenko}.


\section{The RICE signal path}
\label{sec:sigpath}

The neutrino detection efficiency is determined by a detailed Monte Carlo 
simulation of the RICE experiment, which naturally divides into four 
stages: 1) the incident neutrino flux and the neutrino-nucleon 
interactions,  2) generation of the subsequent electromagnetic shower and 
the radio signal from that shower,  3) propagation of the radio signal 
through the ice, resulting in an electric field $\vec{E}_i(t)$ at the 
input to the $i^{th}$ antenna, and 4) response of each antenna, and 
conversion to an input voltage $V_i(t)$ for each channel received by the 
DAQ hardware. 

The primary result of the Monte Carlo is an energy-dependent effective 
volume for detection of electron neutrinos which undergo charged current 
scattering from nucleons in ice and so create an electromagnetic shower. 
For a given model spectrum and integrated livetime, the expected number 
of detections is then readily determined.  We compare this with the 
observed number of detections, and obtain an upper limit to the 
normalization of the model spectrum.


\subsection{Shower Production Rate}
\label{sec:showerrate}

The simulation assumes electron neutrinos of energy $E_\nu$ are incident 
isotropically from the {\it upper hemisphere}.  The upward neutrino flux 
is negligible due to the high opacity of the Earth for $E_\nu>$1 
PeV\cite{GandhiQRS96}. Events are distributed uniformly over a 
cylindrical volume with a depth and radius large enough to contain the 
most distant detectable events for a given neutrino energy. The design 
energy and angular range of the RICE search, as well as the neutrino 
flavor, all differ from those of lower energy cosmic neutrino 
experiments, which primarily search for upward-going muons.

The probability of a charged current interaction within the Antarctic ice 
is determined using the cross-sections of Gandhi et 
al.\cite{GandhiQRS98}. Other extrapolations of Standard Model 
cross-sections to the high energy range differ by as much as 20\% at 10 
EeV\cite{FrichterMR,GRV98,GluckKR,KwiecinskiMS}. For oxygen, the 
per-nucleon cross-section is reduced due to nuclear structure 
effects\cite{NMC,E665} by 12\%(1 PeV) to 24\%(100 
EeV)\cite{PenaPZ,DuttaRSS}.  The electromagnetic shower energy is related 
to the neutrino energy by the inelasticity, $E_s = (1-y) E_\nu$. We 
select $y$ randomly from $d\sigma/dy$, also taken from 
Ref.~\cite{GandhiQRS98}. The mean value of $y$ drops from 26\% to 20\% as 
the neutrino energy increases from 1 PeV to 100 EeV. 

RICE sensitivity to hadronic showers is discussed in Section 
\ref{sec:disc}. Hadronic showers may be produced by either charged or 
neutral current interactions. Neutral current cross-sections were also 
taken from Gandhi et al., and similarly modified for nuclear effects. At 
energies above a PeV, the cross-sections are dominated by sea-quarks and 
so, a) the $\bar\nu$ cross-sections are nearly equal to the $\nu$ 
cross-sections, b) the neutral current cross-sections are nearly 0.5 
times the charged current cross-sections, and c) the $d\sigma/dy$ 
distributions are nearly identical for all interaction types.

The energy in hadronic showers evolves into an electromagnetic shower as 
$\pi^0$ mesons in the shower decay to two photons. This process is 
suppressed for $E_{\pi^0} \simgt 6$ PeV, because the rate for hadronic 
interactions of $\pi^0$'s exceeds the time dilated decay rate. As a 
result, conversion of hadronic energy to electromagnetic energy is 
postponed until average particle energies are below a few PeV, with the 
consequence that hadronic showers do not suffer greatly from LPM 
effects\cite{AlvarezZ98}. In our later discussion we include hadronic 
showers as electromagnetic showers with energy $E_s = y E_\nu$ and 
discount any LPM effects.


\subsection{Expected Signal Strength due to an Electromagnetic Shower}
\label{sec:empulse}

The coherent emission of radio-frequency Cherenkov radiation from 
cascades has been discussed by several 
authors\cite{ZasHS,FrichterRM,RalstonM,AlvarezZ97,AlvarezZ98,AlvarezVZ,BuniyR,RazzaqueSBMRS,AlvarezMVZ,RazzaqueSBMRSa}.
Laboratory measurements directly probe the coherence and have verified a 
linear scaling of the field strength with shower energy\cite{Saltzberg}, 
but a precise determination of signal strength for showers developing 
in ice still requires numerical simulation.

The most recent papers compare radiation from showers produced using the 
GEANT particle physics and detector Monte Carlo package with codes that 
are derivatives of the pioneering work of Zas, Halzen and 
Stanev\cite{ZasHS} (ZHS). Razzaque et al.\cite{RazzaqueSBMRS} used GEANT 
3.21 and found that emission from showers with energies of 0.1-1 TeV was 
typically 30\% lower than in ZHS at frequencies $f < 500$~MHz, the regime 
for fully coherent emission from the shower. Subsequently, Alvarez-Mu\~niz 
et al.\cite{AlvarezMVZ} showed that the low result from GEANT 3.21 was 
due to an approximation made in the tracking of low energy 
electrons. Utilizing an optional tracking algorithm, or switching to 
GEANT 4, produced results only 10\% smaller than the ZHS code. This result 
has been confirmed by Razzaque et al.\cite{RazzaqueSBMRSa}, who also show 
that other characteristics of the radio pulse are not greatly affected by the 
change in GEANT configuration.

In addition to overall normalization, the frequency spectrum (and phase) 
convolved with the experimental bandwidth determines the detected signal 
strength. For both the GEANT and ZHS codes the power spectrum increases 
with frequency until $f \sim 1$~GHz. At higher frequency the ZHS code 
produces a spectrum which flattens and then decreases in the multi-GHz 
range\cite{ZasHS,AlvarezVZ}, whereas the spectrum from the GEANT code 
remains flat for $f > 1$~GHz\cite{RazzaqueSBMRS}. These differences are 
still being studied, but are relatively unimportant for RICE, where 
propagation in ice, antenna response, cable losses, and noise filters 
reduce the effective bandwidth to the range $\Delta f \approx 
200-500$~MHz\cite{Kravchenko}.

The angular radiation pattern is mainly determined by the length of the 
shower, for which the GEANT and ZHS codes are in agreement at 
energies of 100 GeV\cite{RazzaqueSBMRS}. However, the peak of the RICE 
energy response is some 6 orders of magnitude higher, so an extrapolation 
from low- to high-energy must be performed. Accordingly, we include the 
modest lengthening of the shower when extrapolating from 1 TeV to 1 
PeV\cite{Greisen}, and for electromagnetic shower energies above one PeV 
the radiation pattern is adjusted to include LPM 
effects\cite{RalstonM,AlvarezZ97}. 

Given these considerations, for our Monte Carlo simulation we use a radio 
pulse with spectral form taken from the original ZHS paper, but with a 
normalization given by the most recent GEANT studies and an angular 
pattern which includes shower evolution and LPM effects. We estimate that 
the potential systematic error in this procedure is of order 10\% for an 
average shower. We do not model shower-to-shower fluctuations, nor do we 
include the near zone corrections to the radiation pattern\cite{BuniyR} 
that may be be important for impact parameters of a few hundred meters.

For our later discussions, we approximate hadronic showers as 
electromagnetic showers of the same energy, but with LPM effects turned 
off. This approximation ignores modest differences in the longitudinal 
development of the shower due to its hadronic nature. In the case of 
combined electron induced and hadronic showers, due to the different 
longitudinal development, the two pulses may not add coherently at all 
observation angles. Thus, we approximate the two pulses as distinct 
objects and evaluate their detectability separately. This procedure will 
sometimes result in double counting an event, and in other cases will 
miss events where neither shower would trigger RICE by itself, but the 
combined showers would.


\subsection{Propagation through Polar Ice}
\label{sec:iceprop}

Radio propagation in ice\cite{BogorodskyBG} is described by a complex 
dielectric constant $\epsilon$,  which in principle is a function of the 
density, temperature, composition, and crystalline structure of the ice, 
as well as the frequency and polarization of the radio wave. The real 
part of $\epsilon$ defines the index of refraction $n_{ice}$ and depends 
primarily on density, while the complex part of $\epsilon$ is related to 
absorption or, equivalently, the attenuation of field strength. We use an 
index of refraction $n_{ice} = 1 + 0.84 \rho$, where $\rho(z)$ is the ice 
density in gm/cm$^3$, which yields $n_{ice} = 1.3$ near the surface and a 
deep ice value of 1.78. The ice is assumed to be free of discontinuities. 
The attenuation length is taken to be frequency and temperature 
dependent. (Note: we work with the field attenuation length, which is 
twice the power absorption length.)

We have searched for ``echoes" in transmitter data indicative of 
reflections from interfaces in the ice or from the surface, but have 
found no evidence for such an effect. Surface events do show evidence for 
``double pulses'', but the time separation is typically independent of 
receiver channel, showing that the double pulses are properties of the 
source and not features of the propagation. Data from nadir looking 
synthetic aperture radar suggest that variations of $n_{ice}$ at internal 
interfaces in the ice are quite weak, generating reflected waves down by 
20-90 dB relative to the wave transmitted through the 
interface\cite{sar}. Non-stratified variations in $n_{ice}$ may lead to 
incoherent scattering, a loss of signal strength, and a delayed clutter 
of scattered waves. We see no evidence for such effects in our buried 
transmitter data.

\begin{figure}[htpb]
\centerline{\includegraphics[width=10cm]{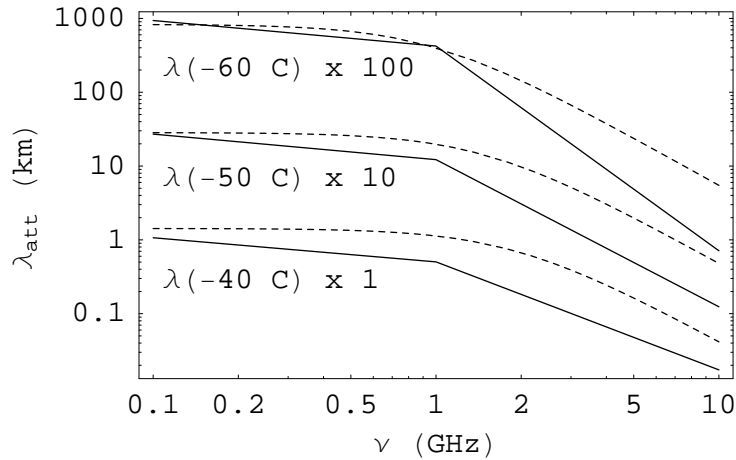}}
\caption{
Models for attenuation length, $\lambda_{att}(\nu,T)$ at three 
temperatures. Solid curves are for a parameterization due to 
Provorov\cite{Provorov}. Dashed curves are for a model combining data 
from Matsuoka, Fujita and Mae\cite{MatsuokaFM}, and 
Westphal\cite{Westphal}. The pairs of curves are offset for visibility.
}
\label{fig:latt}
\end{figure}

Figure \ref{fig:latt} shows two models for $\lambda(\nu,T)$, the 
frequency and temperature dependent attenuation length. The current 
analysis makes use of a parameterization due to Provorov\cite{Provorov}. 
We have also constructed a model combining high frequency data ($\nu > 5$ 
GHz) taken with pure polycrystaline ice in the 
laboratory\cite{MatsuokaFM}, with low frequency measurements ($\nu < 600$ 
MHz) taken on a sample of glacial ice from Greenland\cite{Westphal}. The 
two models differ by less than a factor of two in the RICE bandpass of 
200-500 MHz. Although there is agreement between these two models, 
other measurements of attenuation at 300 MHz vary significantly. 
Extrapolation of Debye spectra measured for pure ice at lower 
frequencies\cite{Kawada,Tonkonogov} yield attenuation lengths more than 
an order of magnitude longer at -50 C, while data from a sample taken in 
Little America\cite{BogorodskyBG} indicate $\lambda$ a factor of 3 
smaller at -40 C. We speculate that the differences are due to varying 
levels of impurities affecting the number of charge carriers that 
contribute to conduction at low frequency\cite{Fujita,Tonkonogov}. The 
ice at South Pole should be relatively impurity-free due to its remote 
location and isolation from wind-borne sea salts. Attempts to measure the 
attenuation length directly using the RICE array were unsuccessful, 
primarily because the attenuation length in the RICE bandpass exceeds one 
km, far beyond the size of the array and its buried transmitter-receiver 
pairs. 
%


\subsection{From Antenna to Trigger}
\label{sec:atot}

Knowing the amplitude and timing response of the RICE receiver array is 
crucial to determining the sensitivity for neutrino events, as well as 
the ability to reconstruct events and reject spurious backgrounds. Both 
these calibrations are most easily analyzed in the frequency domain - 
transforming back to the time domain only as the signal enters the DAQ 
(trigger logic and oscilloscopes). The signal incident to the DAQ is 
given by
\begin{equation}
V_o(\nu) = T(\nu) V_i(\nu) \,,
\label{eq:vivo}
\end{equation}
where $T$ is the transfer function and $V_i$ is the voltage generated on 
the antenna. For a dipole antenna with effective height $\vec{h}$, the 
input voltage is given $V_i = \vec{h} \cdot \vec{E}(\nu)$, where 
$\vec{E}(\nu)$ is the electric field vector of the Cherenkov signal. 
The transfer function to pass signals through to the DAQ is given by
\begin{equation}
T(\nu) = A(\nu) \frac{Z_{ant}}{Z_{ant}+Z_{cable}} \,,
\end{equation}
where $A(\nu) = A_1 C A_2 F/\sqrt{2}$ is a product of terms to account 
for the amplifiers ($A_1$ and $A_2$), cable losses (C), the 200 MHz 
filter (F) and the splitter, $Z_{cable}=50\Omega$, and $Z_{ant}(\nu)$ is 
the frequency dependent antenna impedance. 

The University of Kansas Antenna Testing Range has been used both to 
measure the effective height for RICE antennas and to confirm a basic 
dipole antenna pattern. Amplifier gains are measured {\it in situ}. Cable 
losses and the effects of filters and splitters are measured in the lab. 
The full frequency response is then tested using continuous wave 
transmissions from the in-ice transmitters. The model reproduces the 
shape and normalization of data, with errors no greater than $\pm$3 dB in 
voltage, antenna-by-antenna, across the frequency 
spectrum\cite{Kravchenko}. 

\begin{figure}[htpb]
\centerline{\includegraphics[width=10cm]{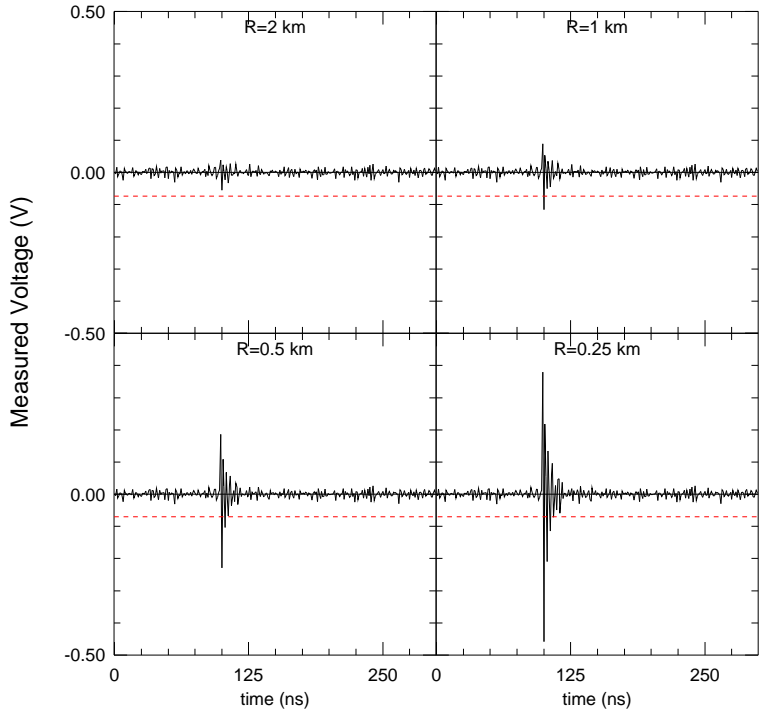}}
\caption{
Voltage as a function of time expected for a 10 PeV cascade at different 
distances from a typical RICE receiver. The receiver is positioned on the 
Cherenkov cone with its dipole axis aligned along the electric field. The 
noise background is sampled from actual data. The horizontal dashed line 
is the `$5-\sigma$' threshold used in analysis. 
}
\label{fig:Vt_embedded}
\end{figure}

After the signal arrives at the DAQ, the Monte Carlo simulates the RICE 
trigger by transforming the signal back to the time domain, comparing to 
the relevant discriminator thresholds and checking for a 4-fold 
coincidence within 1.2 $\mu$sec. Sample results are shown in Figure 
\ref{fig:Vt_embedded}. 

Details of the antenna array and DAQ also influence the timing accuracy, 
critical for event reconstruction. A fairly straight forward survey of 
signal delays within the cables and DAQ electronics indicates that these 
contribute $< 5$ns to the timing uncertainty\cite{Kravchenko}. 
Uncertainty in antenna positions is more important, $\sim 4$ m or 20 ns. 
A more subtle issue is consistently determining an arrival time from the 
pulse waveform. In Figure \ref{fig:Vt_embedded} one observes a 
characteristic ringing of our antennas at approximately 300 MHz, and 
damped with a time constant of order 10 ns. Still, the leading edge of 
the pulse can be resolved to a few ns on an event-by-event basis. Pulse 
shapes for noise events originating from the surface are not as clean or 
reproducible, particularly for high-amplitude signals which saturate the 
amplifiers, and an arrival time uncertainty of 50 ns is more typical. For 
tests involving the in-ice transmitters, the pulses are reproducible, but 
the impulse is considerably broader than for a neutrino event. We expect 
that strong neutrino pulses can be consistently timed to $<2$ ns, but 
weaker pulses give uncertainties of order 10 ns. 


\subsection{Effective Volume}
\label{sec:veff}

We define the effective volume of the array for shower detection by
\begin{equation}
V_{eff}(E_s) = \frac{1}{2\pi\tau} \int \epsilon(E_s,\vec{r},\hat{n},t) 
d\vec{r} d\Omega dt \,,
\label{eq:veff}
\end{equation}
where $\epsilon(E_s,\vec{r},\hat{n},t)$ is the efficiency for detection 
of an electromagnetic shower of energy $E_s$, at position $\vec{r}$, and 
direction $\hat{n}$. We normalize $V_{eff}$ to $2\pi$~sr to account for 
the solid angle of the incident beam and average over the exposure time 
$\tau$ since $\epsilon$ depends on adjustable DAQ settings.

\begin{figure}[htpb]
\centerline{\includegraphics[width=10cm]{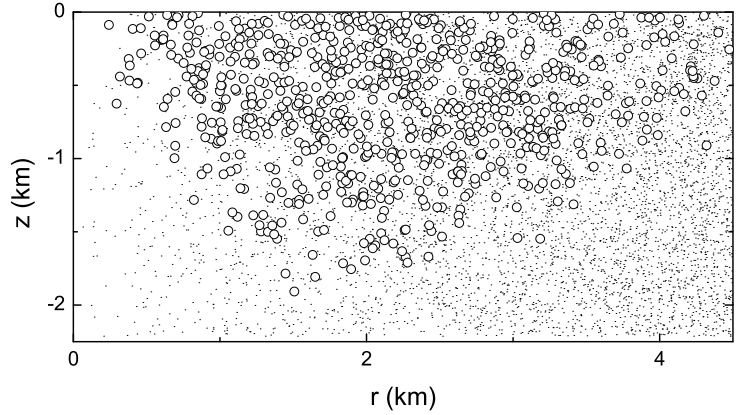}}
\caption{
Radial and depth distribution for 60000 Monte Carlo events at shower 
energy of 1 EeV. The Monte Carlo exposure is even in depth, unit area and 
solid angle above the horizon. Black dots show event vertices for 10\% of 
the simulated events. Open circles indicate the positions of simulated 
events which would result in RICE triggers. 
}
\label{fig:zvsr}
\end{figure}

Figure \ref{fig:zvsr} illustrates $\epsilon$ for showers with $E_s = 
1$~EeV.  At small radial distances, downward showers initiated below the 
array are not detected since the Cherenkov cone does not intercept the 
array. At modest distances, the efficiency is limited by the width of the 
Cherenkov cone. At large distances, attenuation in the ice limits the 
horizon of the detector. Most of the efficiency is in the upper km of 
ice, since the ice warms with depth and the effects of attenuation 
increase. Averaged over depth, the efficiency for this exposure is a 
maximum of about 3\% from 1-2 km, and is less than 0.3\% for $r > 4$~km.

\begin{figure}[htpb]
\begin{center}
\includegraphics[width=12cm]{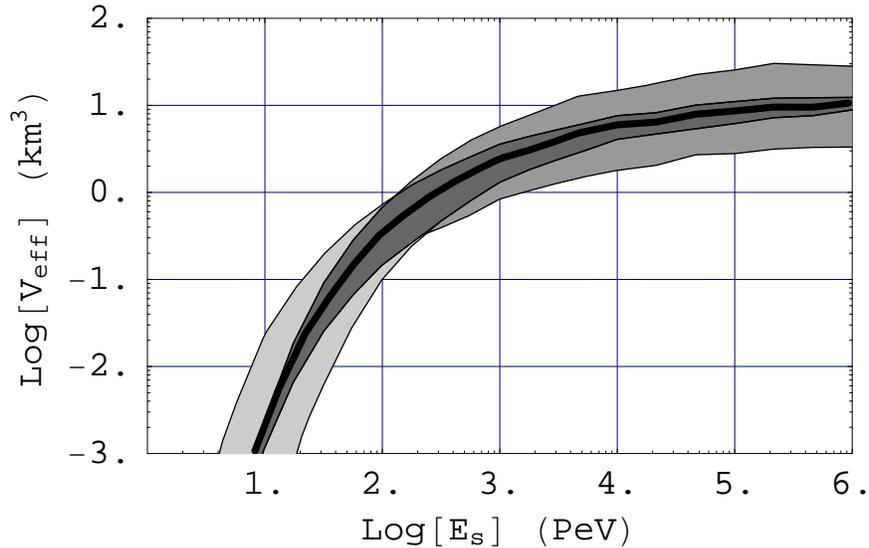}
\end{center}
\caption{
RICE effective volume (Eq. \ref{eq:veff}), as a function of shower 
energy, for the August 2000 configuration. The nominal result corresponds 
to the bold curve. The region in medium gray spans variations in response 
due to variation in the attenuation length by factors of (0.5-2.0). The 
region in light gray shows the range due to changes in signal strength by 
(0.5-2.0). The region shaded in dark gray is within the range of both 
variations.
}
\label{fig:eff_volume}
\end{figure}

The effective volume averaged over the August 2000 exposure is shown as 
the bold curve in Figure \ref{fig:eff_volume}. For $E_s = 300$ PeV, 
$V_{eff}$ is $\sim$ 1 km$^3$. 
For energies below about 50 PeV, $V_{eff} \sim E^3$\cite{FrichterRM}, 
reflecting the nearly linear scaling of the signal strength with shower 
energy. Above 50 PeV three effects become important as the range for 
signal detection increases: a) the target volume becomes more and more 
disk-like, so that $V_{eff} \sim E^2$, b) eventually attenuation 
dominates so that $V_{eff} \sim \ln(E)^2$, and c) LPM effects narrow the 
thickness of the Cherenkov cone, further reducing $V_{eff}$.

Figure \ref{fig:eff_volume} also illustrates two complementary 
contributions to uncertainty in the analysis. Any variation in the 
attenuation length is most important at high energies, where the array is 
probing distances well over 1 km. Variations in signal strength, however, 
are most important for energies below 100 PeV. The latter regime is also 
most sensitive to calibration of receiver sensitivities and changes in 
the discriminator threshold or trigger conditions. For example, changing 
the trigger requirement to a 5-fold multiplicity, increasing the 
discriminator threshold by 25\%, or decreasing the signal strength by a 
similar amount, all have approximately the same effect on $V_{eff}$.


\section{Event Reconstruction}
\label{sec:recon}


Event reconstruction is performed in two stages: 1) an event vertex is 
determined based on pulse timing and 2) neutrino direction and shower 
energy are determined from fitting a Cherenkov cone to the signal 
amplitudes in channels registering hits.

\begin{figure}[t]
\begin{center}
\includegraphics[width=12cm]{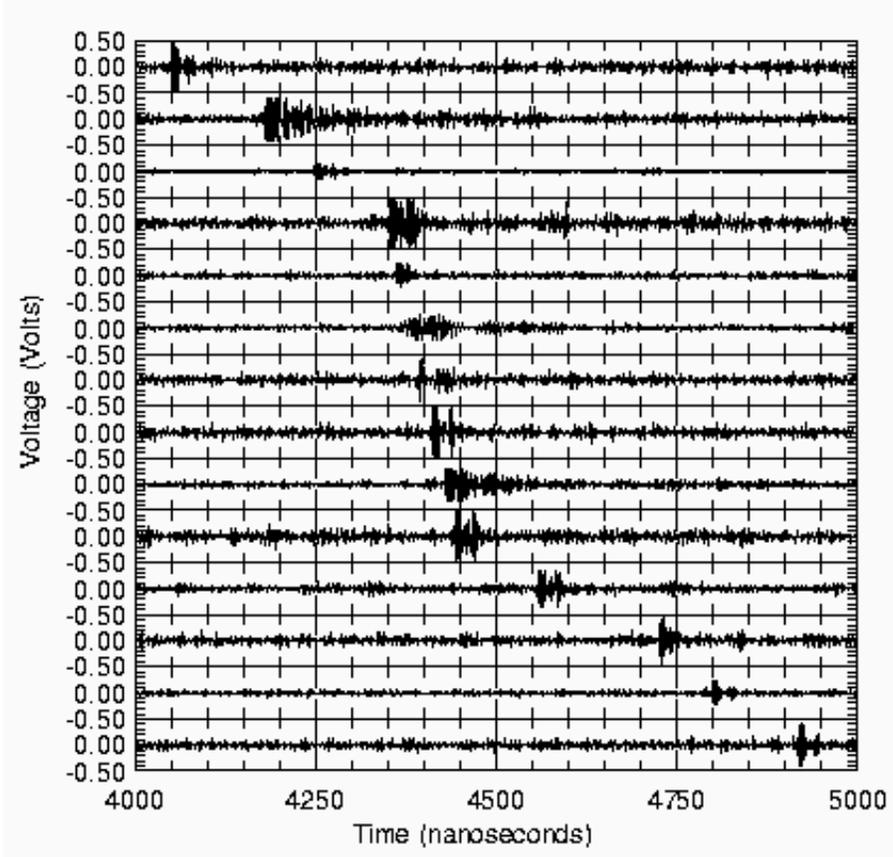}
\end{center}
\caption{
Voltage traces for a calibration event in which a 5 ns square pulse is 
sent down cable to an in-ice transmitter and broadcast to the array. 
Cable losses and bandwidth limitations of the transmitting antenna 
broaden the time structure of the pulse significantly. Receiver channels 
are ordered (from top to bottom) in order of the modeled time delay 
between the pulse transmission and reception in the DAQ. There are only 
14 traces in the figure as two oscilloscope channels were dedicated to 
triggering in this test. 
}
\label{fig:txrxtest}
\end{figure}

Techniques for vertex reconstruction were developed using short duration 
in-ice transmitter pulses broadcast to the receiver 
array\cite{Kravchenko}. Given measured time differences $\delta t_{ij}$ 
between all pairs $(i,j)$ of hit antennas, we perform a $\chi^2$ 
minimization to reconstruct the source location. A typical transmitter 
calibration event is displayed in Figure \ref{fig:txrxtest}. The relevant 
time scales are evident. Compared to the dimensions of the array, the 
pulses are well resolved. Although the pulses themselves extend over 
$\sim 50$ ns, the pulse shapes are repeatable with arrival times 
determined to better than 2 ns rms. Combining uncertainties due to in-ice 
propagation, surveyed positions of the receivers, propagation delays 
through the array cables and electronics, and pulse time resolution, we 
infer a precision for determining transmitter positions of better than 
5 m, corresponding to a time resolution of $\sim 25$ ns.
%


The same techniques are used for analysis of live data and to 
discriminate between potential signal events and various backgrounds.
The vertex reconstruction is underdetermined for three or fewer hits, so we 
require four hits to form a trigger.  With four hits, a geometrical 
reconstruction typically admits either a unique solution or no solution 
at all. Thus, for a point source event $\chi^2 \rightarrow 0$, whereas 
for any finite $\chi^2$ a valid vertex reconstruction is not possible. 
With five or more hits, a valid reconstruction should yield a 
$\chi^2/$d.o.f. of order 1. 

To discriminate against surface noise, a timing uncertainty $\delta t 
\sim 50$ ns is appropriate due to the uncertain nature of the pulse 
shapes. For vertices within the array boundaries or nearby on the 
surface, we reconstruct positions to roughly 10 m accuracy - sufficient to 
distinguish surface events from in-ice neutrino candidates.

For reconstruction of neutrino event candidates, a smaller $\delta t \sim 
25$ ns is dominated by uncertainty in receiver positions. More generally, 
for events within a km of the array we obtain vertex location resolutions 
of $\sim$0.1 R, where R is the distance from the vertex to array center. 
Radial distance is not well determined beyond 1 km, where the timing 
accuracy does not allow one to distinguish between a spherical wavefront 
and a plane wave. 

We utilize the relative signal amplitudes amongst the hit antennas to 
determine the section of the Cherenkov cone detected and, hence, the 
direction of the incident neutrino. The difference between true and 
reconstructed sky position is less than $\sim$10 degrees for half of the 
simulated neutrino events. Similarly, the energy resolution for  showers 
at 100 PeV is $\Delta E/E \sim 0.5$ for those cases where the Cherenkov 
geometry is well-determined\cite{Kravchenko}.


\section{Backgrounds and filters}
\label{sec:back}

The radio environment at South Pole includes both a thermal background 
and a variety of anthropogenic effects\cite{Frichter99}. The 200 MHz high 
pass filter eliminates low frequency backgrounds before signals reach the 
DAQ. Data-taking is inhibited while the station satellite link (303 MHz) 
is operating. Still, the raw RICE trigger rate is dominated by 
electronics noise generated on the surface, mostly in or nearby MAPO. 
These events can be readily eliminated by a sequence of criteria based on 
the reconstructed vertex location and the quality of the event pulse 
shape, as detailed in the filter and analysis chain given below.

Thermal noise fluctuations represent an irreducible background even after 
all impulsive human activity has been eliminated. Each receiver is 
characterized by a measured effective noise temperature $T_n \sim 
250-300$ K, which includes the ambient radiation field, the physical 
temperature of the antenna, and the noise generated by the in-ice 
amplifier. The thermal noise contribution to $V_i$ in Eq. \ref{eq:vivo} 
is given by 
\begin{equation}
V_{rms} = (k_B Z_{cable} T_n \Delta\nu)^{1/2} \,,
\end{equation}
where the effective bandwidth $\Delta\nu$ is limited by cable losses, the 
filter, and by either the oscilloscope or the discriminator.  The primary 
mechanism for reducing thermal noise is to require an $n-{\rm hit} \times 
m-\sigma$ trigger, within the 1.2 $\mu$s coincidence window, where 
$\sigma$ is the corresponding noise contribution to the voltage as seen 
by a particular DAQ channel.  With 4-hits required for event 
reconstruction, a $5-\sigma$ threshold gives an expected rate of order 
$10^{-5}$ Hz. As a practical matter, the discriminator threshold is set 
at a level which corresponds to roughly $4-\sigma$ as recorded by the 
oscilloscope; however, the bandwidth for the discriminator is somewhat 
less than that of the oscilloscope and thermal fluctuations do not 
strongly influence the trigger rate. 

\begin{figure}[htpb]
\centerline{\includegraphics[width=10cm]{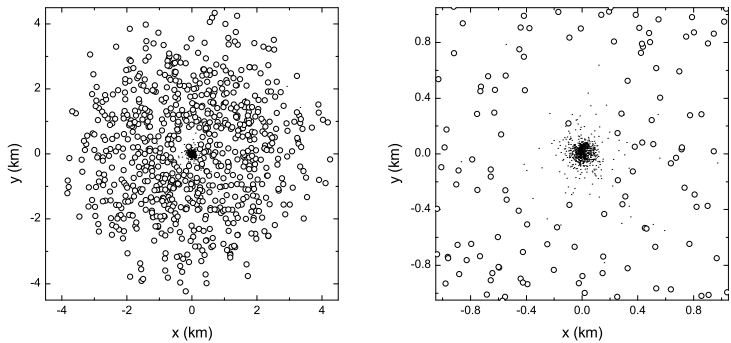}}
\caption{
Lateral distribution of Monte Carlo events for 1 EeV showers (open 
circles) as compared to the vertex position constructed for events due to 
4 uncorrelated noise hits (dots). The latter predominantly lie within the 
array boundaries, whereas true neutrino events do not. The deficit of 
neutrino events in the inner $\sim 0.5$~km reflects the same geometric 
exclusion illustrated in Fig. \ref{fig:zvsr}. Left panel: all events. 
Right panel: inner kilometer. There are 778 points in each sample. 
}
\label{fig:xyvertices}
\end{figure}

During event analysis we apply various cuts based on the waveform data to 
eliminate any remaining ``thermal" triggers. In addition to the $n-{\rm 
hit} \times m-\sigma$ criterion, one can discriminate against such events 
on the basis of vertex location, as shown in Figure \ref{fig:xyvertices}. 
Events where uncorrelated noise results in the production of four random 
hits within 1.2 $\mu$s tend to reconstruct within the array, whereas most 
of our source models are dominated by events at a distance of several km. 
Additionally, events from uncorrelated noise typically do not give good 
fits to a Cherenkov cone. 

Given the high threshold, RICE does not have a naturally occurring source of 
impulsive backgrounds that would obscure the neutrino induced signals. For 
example, catastrophic $dE/dX$ from ultra high energy muons could mimic a 
neutrino event, but the rate for this process is expected to be less 
than 0.1 event per yr\cite{Frichter98}, and generically is smaller than the 
neutrino detection rates. Although this implies that RICE is essentially a 
signal limited experiment, it also means that there is no naturally 
occuring source which may be used to calibrate the experiment.

The full filter and analysis chain used to separate candidate neutrino 
events from impulsive and thermal backgrounds consists of the following 
steps.

\begin{itemize}


\item {\it On-line filter}: The majority ($>99\%$) of events originating 
from the surface are eliminated by a software algorithm based on pairwise 
timing correlations. The algorithm, which uses the TDC times recorded for 
each channel containing a hit, takes $\sim 25$ms to execute. Events which 
would vertex in the ice pass with high efficiency. Events which pass this 
software filter have waveforms read out from the digital oscilloscopes 
for further analysis.

\item {\it Elimination of CW frequencies}: Waveforms are transformed to the
frequency domain, scanned for lines, filtered, and transformed back to
the time domain.

\item {\it Scanning for 5$\sigma$ fluctuations}: The first $\mu$s of 
each waveform is used to evaluate the ambient background noise level for 
that channel. As these are consistent with in-ice thermal noise, it forms 
a physical basis for calibrating the signal strength in hit channels. The 
remainder of the waveform is scanned for well separated 5$\sigma$ 
excursions. A list of pulses for each waveform is generated.

\item {\it Double-pulse elimination}: Many surface events contain 
channels with multiple pulses, which may allow such events to pass the 
on-line filter. Events with multiple pulses in a single channel are 
rejected.  

\item {\it 4-fold 5$\sigma$ criteria}: Four hits must occur within 1.2 
$\mu$s. 

\item {\it Vertex reconstruction}: We require that the hits admit to an 
acceptably well-fit vertex. With four hits, this is a simple yes/no 
decision based on the reconstruction $\chi^2$. For events with five or 
more hits, if $\chi^2$ is good the event is accepted. If $\chi^2$ is 
large, the vertex reconstruction is recalculated after elimination of any 
one hit, allowing for the possiblity that a chance noise hit is captured 
at the same time as a neutrino event. With 8000 samples and a $5-\sigma$ 
threshold, we estimate a chance occupancy of $\sim 4 \times 10^{-3}$ per 
channel, or somewhat less than 7\% probability of contamination per 
event. If the new $\chi^2$ is also large, the process is iterated until 
either $\chi^2$ is acceptable or the event contains less than 4 hits, at 
which point it is rejected.

\item {\it$z < -50$ m} : The reconstructed vertex is required to be at 
least 50 m below the ice surface. 

\item {\it Hand Scanning}: Remaining events are scanned, by eye, for 
patterns or defects. 

\item {\it Event reconstruction}: Each event is tested against the 
hypothesis that the hits are due to a Cherenkov cone radiation pattern 
from the reconstructed vertex. The cone constructions are analyzed by 
hand to ensure that acceptance/rejection is robust.

\item {\it Waveform quality}: The waveforms are compared to simulated 
waveforms (e.g. Figure \ref{fig:Vt_embedded}) to ensure that the time 
structures of the recorded signals are consistent with model predictions.

\end{itemize}

The detector Monte Carlo simulation has been used to test the filter and 
analysis sequence. Simulated neutrino events which contribute to 
$V_{eff}$ pass the on-line filter with $80 \pm 5\%$ efficiency. The remaining 
stages pass simulated events with $85 \pm 5\%$ efficiency. These factors 
are not included in $V_{eff}$ shown in Figure \ref{fig:eff_volume}, but are 
included in subsequent calculations of event rates and limits.


\section{Analysis of August, 2000 Data}
\label{sec:augytwok}

During August 2000,  the RICE experiment was operated in a stable 
configuration for a total of 660.1 hrs. Data-taking was suppressed while 
the South Pole Station satellite link was operational ($\sim 242$ hrs). 
Additional dead time accrued during the execution of the on-line filter 
algorithm ($\sim 35$ hrs), during the readout of events which passed the 
on-line filters ($\sim$ 27 hrs), and during readout of unbiased 
events and a sample of vetoed events taken for monitoring purposes 
($\sim$ 23 hr). All told, the exposure was 333.3 hrs of livetime. 

Table \ref{tab:cuts} summarizes the number of hardware triggers and the 
success with which they pass our various filter criteria.
Evidently, the trigger rate is dominated by surface-generated transient 
noise backgrounds. The on-line filter successfully removes 99.8\% of these 
in real time. Of the remainder, only 0.2\% pass automatic 
scanning procedures, leaving just 22 events to be looked at by hand.
\begin{table}[htpb]
\begin{center}
\begin{tabular}{l|r} \hline
Total triggers & 5,058,976 \\
On line filter & 9,442 \\
$4\times 5\sigma$ cut & 73 \\
double-pulse, $Z<50$m & 22 \\
defect and pattern checks & 1 \\
Cherenkov cone test and waveform quality & 0 \\ \hline
\end{tabular}
\end{center}
\caption{\label{tab:cuts} Summary of analysis of August, 2000 data.}
\end{table}
Of the 22 events scanned by trained physicists, 11 exhibited one of three 
repeated patterns of antenna hits and relative times, indicating a 
man-made source. Another 7 had multiple pulses, but with significant 
activity in the first microsecond of waveform data which may have 
defeated the automated double-pulse filter. Three events, all occurring on 
the same day, had pulses only at the tails of the waveforms, indicative 
of a significant spectral correction for a transient continuous wave 
transmission. One event remained, which was subjected to further analysis.

The pattern of antenna hits failed the test for consistency with a 
Cherenkov-cone originating from the pre-determined vertex: 
$\chi^2$/d.o.f.$>1000$. Upon examination, the event had four hits that 
were nearly coplanar, with two hits on one side of the array and the 
other two nearly opposite, while the event reconstruction puts the vertex 
nearly on the same plane and in the middle of the array. This 
configuration cannot be fit to a Cherenkov cone. The pulseshapes were 
checked for consistency with neutrino generated signals. They were 
somewhat broad in this regard, 50-200 ns depending on the observers 
definition, but significantly wider than the features evident in Figure 
\ref{fig:Vt_embedded}. Finally, the location of the vertex in the center 
of the array is consistent with an interpretation where the multiple hit 
event arises from uncorrelated signals in the individual channels, 
whereas for the model neutrino spectra we considered, the event vertices 
lie well outside the array boundaries (see Figure \ref{fig:xyvertices}). 
Based on these considerations, we conclude that there were no neutrino 
events in the August 2000 data sample.

\begin{figure}[htpb]
\begin{center}
\includegraphics[width=12cm]{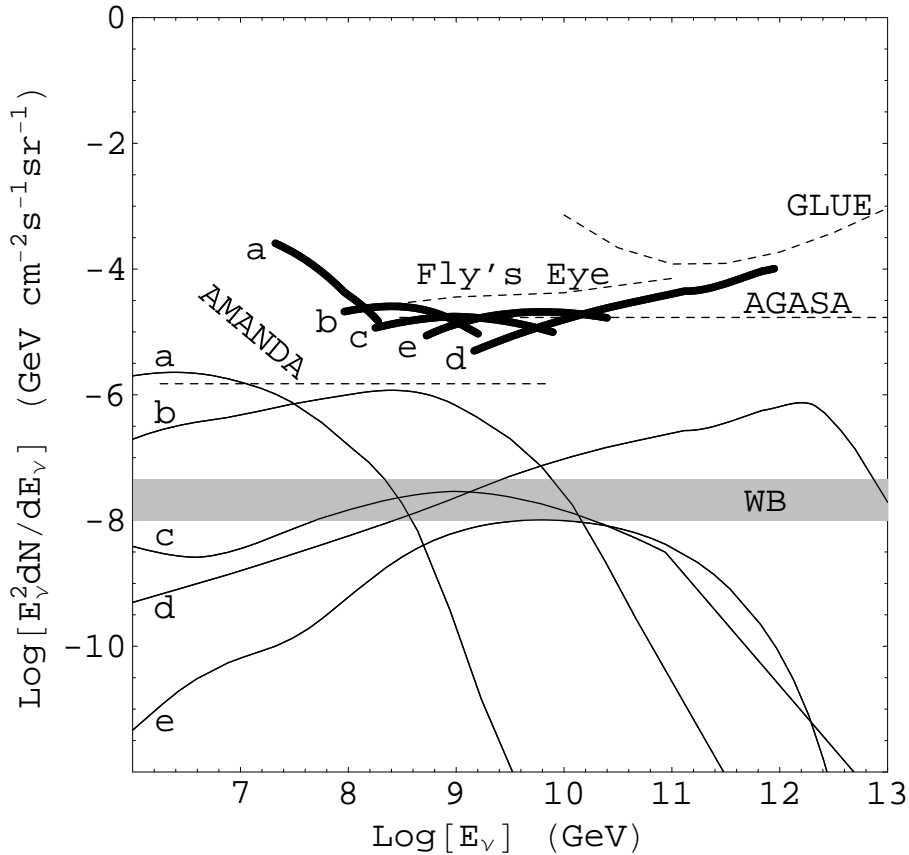}
\end{center}
\caption{
Neutrino flux model predictions (thin) and corresponding RICE upper 
limits (95\% confidence level; thick), as a function of neutrino energy. 
The predictions and RICE limits correspond to 
(a) Stecker \& Salamon\cite{SteckerS} (AGN),
(b) Protheroe\cite{Protheroe} (AGN),
(c) Mannheim\cite{Mannheim} model A (AGN),
(d) Protheroe \& Stanev\cite{ProtheroeS} (topological defect), 
(e) Engel, Seckel \& Stanev\cite{EngelSS} (GZK-model). 
Each segment corresponds to the neutrino energy range responsible for the 
middle 80\% of the event rate. Results from AGASA\cite{agasalim} (assumed 
$E^{-2}$ spectrum for $E_\nu > 3 \times 10^8$ GeV), 
GLUE\cite{GorhamLNSW}, Fly's Eye\cite{flyseyelim} (derived using 
$\sigma_{\nu N}$ from Ref. \cite{GandhiQRS98}), and AMANDA\cite{amandalim} are shown as dashed 
lines. The gray bar (WB) shows the range of upper limits derived by 
Waxman and Bahcall\cite{WaxmanB} under differing assumptions about the 
cosmological evolution of UHE cosmic ray sources.
}
\label{fig:upper_limits}
\end{figure}

Given the total livetime for the August, 2000 dataset, the effective 
volume, and the efficiency for events to pass our filters, we calculate a 
95\% confidence level upper limit on the incident $\nu_e$ flux for 
several theoretical models, as shown in Figure \ref{fig:upper_limits}. 
The limit is derived by first calculating the expected number of events 
during the August 2000 exposure for a given model, and then adjusting the 
model flux by a normalization factor so that the expected number of 
events would be 3.0, the 95\% c.l. bound for zero events and negligible 
background. The limit curves in the Figure then replicate the shape of 
the test spectrum at an amplitude corresponding to the limit. Generally, 
the limits from the current analysis are some 2-3 orders of magnitude 
above model predictions.

\begin{figure}[htpb]
\begin{center}
\includegraphics[width=12cm]{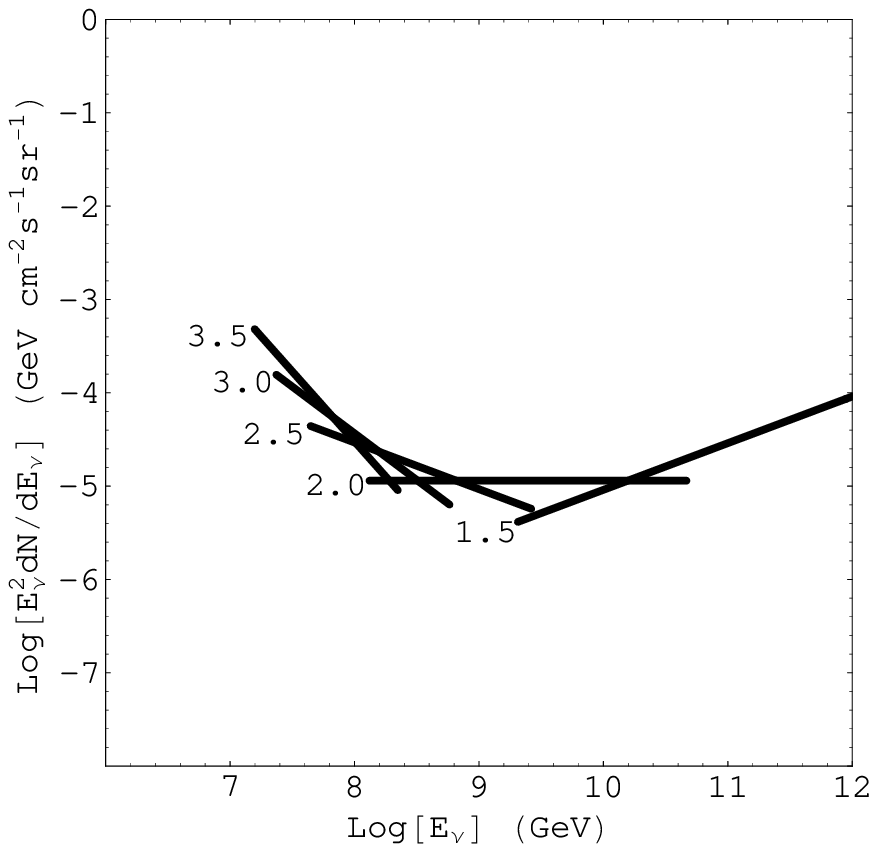}
\end{center}
\caption{
RICE upper limits (95\% confidence level) on neutrino flux assuming a 
$dN/dE\sim E^{-\gamma}$ input spectrum, with the spectral index taking on 
values of $\gamma$=1.5, 2.0, 2.5, 3.0, 3.5.
}
\label{fig:uplim2}
\end{figure}

Alternatively, Figure \ref{fig:uplim2} shows upper limits assuming an
incident power law neutrino energy spectrum: $dN/dE_\nu\sim
E^{-\gamma}$. Again, with no events, the Figure is intended to show
limits on the normalization of the spectrum over the energy range which
gives the greatest contribution to the rate. As the spectral index
increases, the sensitive region is pushed to lower energy.

\section{Discussion}
\label{sec:disc}

The search for ultra high energy neutrinos described in this paper yields 
zero signal candidates. Given the prototype nature of the current RICE 
deployment and the limited exposure used for the present analysis, this 
is not an entirely unexpected result. Accordingly, our discussion is 
focussed on the uncertainties that go into our analysis and on relaxing 
the limitations we imposed on the event types and geometry we consider.

\subsection{Uncertainties}
\label{sec:uncertainties}

Potential systematic uncertainties are summarized in Table 
\ref{tab:sys_sigma}. The Table is organized along the same lines as the 
discussion in Section \ref{sec:sigpath}, where a detailed description of 
input to our Monte Carlo is described. Quoted uncertainties refer to 
particular quantities, not how they propagate through to an uncertainty 
in limits on any specific flux model. 

Uncertainties in neutrino nucleon cross-sections reflect differences 
among available standard model perturbative QCD calculations. Similarly, 
uncertainties in shower simulations and signal generation reflect 
differences between GEANT 4 and ZHS simulations. A discussion of LPM 
issues is given below and leads into the discussion of hadronic showers 
and other event types. A complete understanding of attenuation in polar 
ice is still lacking. The rough factor of two quoted in the table is 
based on the considerations discussed in Section \ref{sec:iceprop}. The 3 
dB uncertainty in channel response is taken from Ref.~\cite{Kravchenko}, 
as are estimates of timing resolution and our ability to reconstruct 
vertex location. Uncertainties for the on-line filter and off-line 
analysis efficiencies are based on Monte Carlo studies. 

\begin{table}
\begin{center}
\begin{tabular}{|c|c|c|}

\hline
Group 
	& Item 
		& $\pm$ \\

\hline
Event Generator 
(Sec. \ref{sec:showerrate})
	& $\sigma_{\nu N}$ 
		& $20\%$  \\

  	& $(1-y)$ 
		& $< 5\%$ \\

\hline
EM pulse
	& shower simulation
		& $10\%$\footnote{Pulse amplitude}  \\
(Sec. \ref{sec:empulse})
	& spectrum
		& $10\%$\footnote{For RICE bandpass of 200-500 MHz} \\
	& extrapolation: TeV $\rightarrow$ PeV 
		& $5\%$ \\
	& LPM effects
		& See text. \\

\hline
Propagation
	& attenuation $\nu < 1 $ GHz 
		& +0.3/-0.15 dB/100 m\footnote{This corresponds  
		to a factor of two uncertainty in $\lambda_{att}$.}\\
(Sec. \ref{sec:iceprop})
	& attenuation $\nu > 1$ GHz 
		& \footnote{Not critical given bandwidth 
		limitations of antennas, cables, amps, and DAQ elements.} \\

  	& index of refraction 
		& $< 1\%$ \\

  	& reflections 
		& small \\

\hline
Antenna and DAQ 
	& overall channel response 
		& 3 dB\\
(Sec. \ref{sec:atot})
	& overall channel timing 
		& 25/50 ns\footnote{25 ns for neutrino events. 50 ns for 
surface generated noise.} \\

  	& on-line filter efficiency
		& $5\%$  \\

  	& dead time 
		& $<5\%$ \\

\hline
Analysis 
	& $V_{eff}$ 
		& 8\%\footnote{Statistical: the Monte Carlo runs 
		produce roughly 150 detections in each $E_s$ bin.} \\
(Secs. \ref{sec:veff}, \ref{sec:recon})
	& nearby vertex reconstruction 
		& 10 m\footnote{w/ assumed 50 ns pulse timing 
uncertainty, appropriate for noise rejection.}\\

	& far vertex reconstruction 
		& 0.1 R\footnote{w/ assumed 50 ns pulse timing 
uncertainty, appropriate for noise rejection.}\\

	& efficiency of analysis chain
		& $5\%$  \\

\hline
\end{tabular}
\caption{\label{tab:sys_sigma}
Estimates of systematic effects for different components
of the RICE analysis.}
\end{center}
\end{table}

The largest uncertainties concern the attenuation in ice at frequencies 
of order 300 MHz and the channel-to-channel calibration of the receivers, 
as summarized by the gray bands in Figure \ref{fig:eff_volume}. 
Uncertainties in calibration of signal strength or receiver sensitivity 
directly affect the threshold for RICE, but do not greatly alter the 
effective volume for shower energies above 1 EeV. This contrasts with the 
effects of modifying attenuation lengths, which would mostly affect 
sensitivity at high energies. These effects are illustrated by Table 
\ref{tab:sys_event}. The Stecker and Salamon AGN model (a), for example, 
produces events at lower energy, and is seen to be most sensitive to 
changes in signal strength and sensitivity, whereas the topological 
defect (d) and GZK (e) models extend to higher energy and are more 
sensitive to uncertainty in $\lambda_{att}$.
\begin{table}
\begin{center}
\begin{tabular}{c|c|cc|cc}
Model & Nom & 2S & 0.5S & $2\lambda_{att}$ & $0.5\lambda_{att}$ \\
\hline
(a) & 0.0120 & 0.0433 & 0.0023 & 0.0220 & 0.0056 \\
(b) & 0.1384 & 0.2597 & 0.0572 & 0.3062 & 0.0526 \\
(c) & 0.0050 & 0.0079 & 0.0025 & 0.0118 & 0.0017 \\
(d) & 0.0182 & 0.0239 & 0.0123 & 0.0494 & 0.0059 \\
(e) & 0.0015 & 0.0021 & 0.0009 & 0.0037 & 0.0005 \\
\end{tabular}
\caption{\label{tab:sys_event}
Expected number of events during a 333.3 hr exposure for different models 
and assumptions about systematic uncertainties. Models are labeled as in 
Fig. \ref{fig:upper_limits}. ``Nom" refers to the nominal case used for generating 
Fig. \ref{fig:upper_limits}. ``2S" and ``0.5S" reflect the change in 
event rate if the signal strength were doubled or halved. ``2$\lambda_{att}$" and 
``0.5$\lambda_{att}$" show the result due to doubling or halving the 
attenuation length.}
\end{center}
\end{table}

To account for the LPM effect, the RICE Monte Carlo uses the 
parameterization of Ref.~\cite{AlvarezZ97}, which describes a narrowing of 
the Cherenkov cone due to a lengthening of the shower. The radiation 
pattern is based on an average shower profile, whereas fluctuations in 
longitudinal development could lead to ``non-gaussian" tails for the 
angular pattern, which would enhance the effective volume. In addition, 
the parameterization is developed in the Fraunhoffer regime, valid for $R 
\gg l^2/\lambda$, where $R$ is the distance to the observer, $l$ is the 
shower length, and $\lambda$ is the wavelength of the radiation. As $l$ 
increases with the LPM effect, a more accurate treatment will lead to a 
broader radiation pattern\cite{BuniyR,AlvarezVZ}, which may enhance the 
effective volume for high energy showers. 

\subsection{Additional event types and hadronic showers}
\label{sec:hadronic}

We have restricted our main analysis to electromagnetic showers produced 
in charged current events of electron neutrinos. The reason for this is 
two fold: a) simplicity, and b) this process is better studied in the 
literature. Even so, expanding the analysis to include neutral currents 
and other neutrino flavors may be expected to increase the rate of 
neutrino interaction candidates by a factor of 4.5. There is a 50\% 
increase due to including neutral currents and a factor of 3 for 
including all flavors. Additional increases may occur if we include 
upward neutrinos from just below the horizon and extra efficiency for 
$\nu_\tau$ due to regeneration in propagation through the Earth and 
$\tau$ decay in the ice near the RICE array. Table \ref{tab:evtype} 
summarizes restrictions we have placed on our analysis, and the potential 
effect on event rates if the restriction were removed.

\begin{table}
\begin{center}
\begin{tabular}{|c|c|}

\hline
Restriction & Effect of relaxation\footnote{Estimates are made at 
$E_{\nu_e} = 1$ EeV, unless stated, where signal from hadronic shower and 
EM shower are comparable (see Figure \ref{fig:veff_lpm}).} \\

\hline
	$\theta_z > 90 \deg$ 
		& Increase rates 10\%\footnote{The Earth is not totally 
		opaque just below the horizon, so rates increase slightly, 
		even at high energies.} \\ 

	no hadronic shower in $\nu_e$ charged current events 
		& 0-100\%\footnote{depends strongly on energy and 
		geometry.}\\

  	CC events only
		& + 50\% per flavor\\

  	$\nu_e$ events only
		& $\times$ 3 \\

	$\tau$ decay not included 
		& increase by 1 charged current channel for $E < 20$ PeV\\

	$\tau$ regeneration not included 
		& increased event rate for upward neutrinos\\

\hline
\end{tabular}
\caption{\label{tab:evtype}
Summary of event type and geometry considerations in RICE analysis.}
\end{center}
\end{table}

As described in Section \ref{sec:showerrate} the hadronic energy is 
typically a quarter of that available on the leptonic side of an event. 
Most of the energy ends up as electromagnetic energy, primarily through 
$\pi^0$ decay; however, that process is not effective until the average 
pion energy drops below $\sim 6$ PeV, similar to the energy where LPM 
effects start to have a significant effect on the longitudinal evolution 
of the shower. As a result, hadronic showers, although decreased and more 
variable in energy, do not suffer the decrease in solid angle acceptance 
for electromagnetic showers related to the LPM effect. 

\begin{figure}[htpb]
\begin{center}
\includegraphics[width=12cm]{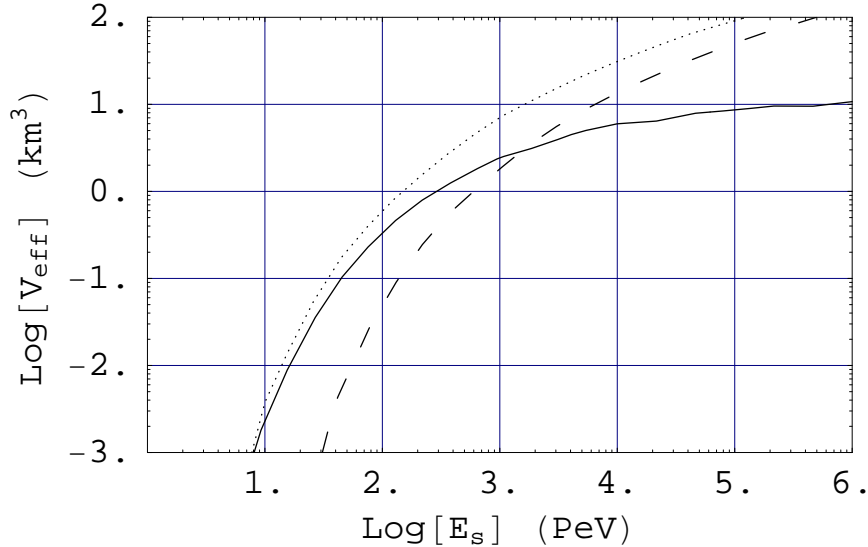}
\end{center}
\caption{
Effective volume with and without the LPM effect. The solid curve is the 
same nominal result with LPM shown in Figure \ref{fig:eff_volume}. The dotted 
curve shows $V_{eff}$ calculated without the LPM effect. The dashed curve shows the 
non-LPM result, but shifted in energy by a factor of 4. This roughly 
mimics the relative response of RICE to hadronic showers (as compared to 
electromagnetic showers from the same energy neutrino), where the average 
inelasticity is $<y> = 0.2$.
}
\label{fig:veff_lpm}
\end{figure}

Given the potential importance of hadronic showers, we have recalculated 
the RICE effective volume under the assumption that LPM effects may be 
ignored - i.e. the signal strength scales linearly with shower energy and 
the angular pattern is energy independent. The result is shown in Figure 
\ref{fig:veff_lpm}. From the Figure, one may infer that the hadronic 
shower provides a stronger radio Cherenkov signal than the 
electromagnetic shower for neutrino energies above $\sim 1$ EeV, for 
typical charged current $\nu_e$ events. 

It is now straight forward to replicate the main analysis, with 
appropriate modifications, to study the potential improvement in 
sensitivity to be gained by considering hadronic showers produced in 
charged and neutral currents of all neutrino flavors. These modifications 
include production of hadronic showers with energy $E_s = y E_\nu$ as 
outlined in Section \ref{sec:showerrate}, and use of the ``no-LPM" 
$V_{eff}$ shown as the dotted curve in Figure \ref{fig:veff_lpm}.

There are three nominal choices to make for the neutrino flavor content, 
which all lead to the same event rates. If there is no mixing, the flavor 
content is determined by the physics of $\pi$-decay, which yields the 
well known relation $\phi_{\nu_\mu} \approx 2 \phi_{\nu_e}$. If there is 
mixing amongst $\nu_\mu$ and $\nu_\tau$ as suggested by the 
SuperKamiokande atmospheric neutrino analysis\cite{superk-atmospheric}, 
but not $\nu_e$, then the event rate of hadronic showers is unchanged, 
since $\nu_\mu$ and $\nu_\tau$ produce the same hadronic showers at these 
energies. If there is full mixing of all flavors, then the rate of EM 
showers is unchanged since the $2:1$ flavor ratio ($\nu_\mu+\nu_\tau : 
\nu_e$) at production is unchanged by mixing. For any of these scenarios, 
Table \ref{tab:had_event} gives event rates including hadronic events. 
For the lower energy AGN models, including hadronic events increases the 
event rate by factors of 2-4. For the GZK (e) and topological defect (d) 
models, the increase is significantly larger - 7 and 16 respectively. 

We have not modeled $\tau$ decay events. To be observed, the $\tau$ would 
have to decay within a km or so of its production site, corresponding to 
$E_\tau \approx 20$~PeV. Thus, charged current $\nu_\tau$ events 
presumably contribute an event rate comparable to the other terms in the 
Table, but which is suppressed at high energy by the probability $P \sim 
\frac{20 {\rm PeV}}{E+20 {\rm PeV}}$ for the $\tau$ to decay in the ice. 
A proper estimate of this term requires a separate Monte Carlo to deal 
with the stochastic nature of $\tau$ decay. Also, since we confine our 
analysis to the downward flux of neutrinos, we do not allow for the 
detection of an upward flux of regenerated $\nu_\tau$\cite{SaltzbergH},
or $\nu_e$ and $\nu_\mu$ produced in $\tau$ decay\cite{BeacomCK}. 
Except near the horizon, these are expected to be of fairly low energy 
compared to the main RICE sensitivity.

\begin{table}
\begin{center}
\begin{tabular}{c|c|cc|c}
Model & $\nu_e$ EM & $\nu_e$ H & $\nu_\mu + \nu_\tau$ H & Total \\
\hline
(a) & 0.0120 & 0.0049 & 0.0098 & 0.0268 \\
(b) & 0.1384 & 0.1096 & 0.2193 & 0.4673 \\
(c) & 0.0050 & 0.0063 & 0.0126 & 0.0238 \\
(d) & 0.0182 & 0.0962 & 0.1923 & 0.3067 \\
(e) & 0.0015 & 0.0029 & 0.0057 & 0.0100 \\
\end{tabular}
\caption{\label{tab:had_event}
Expected number of events during a 333.3 hr exposure for different models, 
including hadronic showers. Models are labeled as in Fig. 
\ref{fig:upper_limits}. ``$\nu_e$ EM" is the nominal event rate shown in 
Fig. \ref{fig:upper_limits} for our main analysis. ``$\nu_e$ H" gives the 
event rate due to the hadronic recoil showers from $\nu_e$ charged 
current and neutral current events. Similarly, ``$\nu_\mu + \nu_\tau$ H" 
gives the event rate due to $\mu$ and $\tau$ neutrinos. For an isoflavor 
model this is just twice the $\nu_e$ H rate. ``Total" gives the sum of 
the three contributions, as appropriate for either an isoflavor model or 
a conventional source model with no mixing and $\phi_{\nu_\mu} \approx 2 
\phi_{\nu_e}$.
}
\end{center}
\end{table}


\section{Summary and Outlook}
\label{sec:summary}

Using data from the RICE antenna array collected during August 2000, we 
have searched for nanosecond radio pulses resulting from neutrino 
interactions in ice. We find no such events, which provides the basis for 
the diffuse flux limits on ultra-high energy electron neutrinos shown in 
Figure \ref{fig:upper_limits} and Figure \ref{fig:uplim2}. This is the 
primary result of this paper.  At the same time, we have demonstrated the 
efficacy of techniques designed to eliminate anthropogenic backgrounds 
and thermal noise.

There are many ways by which either a) the limits we have derived can be 
improved to the level where they seriously constrain models, or b) the 
experiment results in actual UHE neutrino detection. The most obvious of 
these is to increase the exposure and validate sensitivity to hadronic 
showers. Archived data presents a potential increase of more than an 
order of magnitude increase in exposure. Based on the estimates in Table 
\ref{tab:had_event}, including hadronic cascades would increase the event 
rates for typical AGN models by factors of 2-4, and by factors as high as 
10-15 for GZK and topological defect models. Combined, these factors may 
achieve a factor of nearly 100 in sensitivity. This would severely test 
several models - especially the neutrino fluxes associated with top-down 
models proposed to explain the purported excess of cosmic rays with 
energies above the GZK cutoff. It would, however, remain about a decade 
short of testing the Waxman-Bahcall bound\cite{WaxmanB} or probing the 
``guaranteed" flux of neutrinos produced in the GZK process.

To further improve the limits requires improving the acceptance of the 
experiment by reducing the threshold for event detection, or by 
increasing the number of antennas and the area over which they are 
deployed. The main impact of lowering the threshold is to increase the 
sensitivity at modest energies, 1-100 PeV, but, as illustrated in Figure 
\ref{fig:eff_volume}, the gains at higher energy are quite modest. As 
signal attenuation determines the ``horizon" for a single antenna, to 
improve the RICE sensitivity for energies in excess of an EeV requires 
deployment of antennas over a wider area.

There are a number of technological improvements that could dramatically 
improve the sensitivity for energies below 100 PeV, including making use 
of correlated cross-polarized antennas, increasing the bandwidth, making 
use of matched filters to maximize signal to noise, and deploying larger 
numbers of antennas to increase the likelihood that the Cherenkov cone 
from a true neutrino would intersect four or more antennas.

To ensure that the experiment remains signal limited (as opposed to 
background limited) the current noise event rejection must be improved 
without significant loss of efficiency for neutrino events. This should 
be achievable by simple strategies. For example, increasing the number of 
antenna hits required in coincidence from 4 to 5, results in more than a 
hundredfold reduction in thermal noise events, but only a $\sim 15\%$ 
reduction in neutrino induced events. Similarly, a modest increase in the 
restrictions of the on-line veto or the analysis cuts on event 
reconstruction, would reduce impulsive backgrounds significantly without 
significant reduction in neutrino detection efficiency.

Judging by the summary in Table \ref{tab:sys_sigma}, uncertainties 
associated with RICE need to be reduced as well. Most important is the 
need for a better determination of the radio attenuation length in ice. 
We believe the current treatment of $\lambda$ is on the conservative side 
by some 30\%-50\%, which would potentially increase $V_{eff}$ by a factor 
$\sim 2$ for $E_\nu > 1$ EeV. Similarly, we believe the parameterization 
of the LPM effect taken from Ref. \cite{AlvarezZ97} may lead to an 
underestimate of $V_{eff}$. 

Finally, we look forward to the time when true neutrino events are 
detected by RICE, at which point critical issues will turn from acceptance 
and noise rejection to the ability to reconstruct events and identify 
sources and neutrino flavor on an event by event basis. To that end we 
are improving the timing analysis needed for event reconstruction. 
Improving the amplitude calibration is critical here as well, since signal 
strength plays a role in reconstruction of the Cherenkov cone. 

To summarize, we believe the outlook for RICE, and the radio Cherenkov 
technique in general, is quite bright. Analysis of data already in hand 
should significantly improve present limits on $\nu_e$ fluxes. Inclusion 
of other flavors and interaction modes will further constrain models. We 
expect the experiment to remain signal limited, not background limited, 
at least down to levels where even conservative flux models may be 
observed. The ability to reconstruct the vertex and incident neutrino 
direction should permit searches for point sources. Similarly, arrival 
times are sufficiently resolved to conduct coincidence studies between 
RICE events and gamma-ray bursts. Finally, we expect that the technique 
should be extendable to place new limits on ``light" relativistic 
magnetic monopoles\cite{Wick}, as well as neutrino interactions, 
including production of micro black 
holes\cite{AlvarezFHHH,AnchordoquiFGS,KowalskiRT,JainKMPR}.

\begin{acknowledgements}
\input riceack

\end{acknowledgements}


\end{document}

%% file: riceauthor.tex
\author{I. Kravchenko}
\address{Massachusetts Institute of Technology Lab. for
Nuclear Science, Cambridge, MA  02139}
\author{G. M. Frichter}
\address{Florida State University, 
High Energy Physics, Tallahassee FL  32306-4350}
\author{T. Miller\footnote{Current address: Applied Physics Laboratory,
Johns Hopkins University, Laurel MD}, 
L. Piccirillo\footnote{Current address:  Department of Physics and 
Astronomy Cardiff University Cardiff CF24 3YB United Kingdom},
D. Seckel, 
G. M. Spiczak\footnote{Current address:
University of Wisconsin, 
River Falls, WI 54022}}
\address{Bartol Research Institute, U. of Delaware, Newark DE 19716}
\author{J. Adams, S. Seunarine}
\address{Department of Physics and Astronomy,
Private Bag 4800, U. of Canterbury, Christchurch, New Zealand}
\author{C. Allen,
A. Bean,
D. Besson,\footnote{Contact: zedlam@ku.edu, for more information on the
RICE experiment.}
D. J. Box,
R. Buniy,\footnote{Currently at Vanderbilt University, Nashville, TN, 37235}
J. Drees,
D. McKay,
J. Meyers,
L. Perry,
J. Ralston,
S. Razzaque,
D. W. Schmitz\footnote{Currently at Columbia University, NYC, NY  10027}}
\address{University of Kansas Dept. of Physics and Astronomy, Lawrence KS
66045-2151}


%% file: riceack.tex
We gratefully acknowledge the generous logistical support of the AMANDA 
Collaboration (without whom this work would not have been possible), the 
National Science Foundation Office of Polar Programs, the University of 
Kansas General Research Fund and the University of Kansas Research 
Development Fund, the Kansas Center for Advanced Scientific Computing 
kcasc) and the associated NSF-MRI grant for the support of the KU 
Origin2400 supercomputer, the NSF EPSCoR Program, the University of 
Canterbury Marsden Foundation, and the Cottrell Research Corporation. 
Matt Peters of the U. of Texas SOAR group provided essential consultation 
on antenna design, calibration and electrical engineering. The National 
Science Foundation's Research Experience for Undergraduates Program 
provided support for Jeff Allen (Shawnee Mission South High School, 
currently at U. of California, Berkeley), Eben Copple (KU. Physics 
Dept.), Karl Byleen-Higley (Lawrence High School, Lawrence, KS), Adrienne 
Juett (KU Physics Dept., currently at MIT), and James Snow (Eudora High 
School, Eudora, KS), who performed invaluable software assistance and 
antenna and amplifier calibration. Alexey Provorov and Igor Zheleznykh 
(Moscow Institute of Nuclear Research) constructed the TEM horn antennas 
currently used in the surface-noise veto. We also thank the winterovers 
who staffed the experiment during the last three years at the South Pole 
(Xinhua Bai, Allan Baker, Mike Boyce, Marc Hellwig, Matthias Leupold, 
Karl Mueller, Katherine Rawlins, Steffen Richter, and Darryn Schneider.) 
Ryan Dyer helped in deployment during the 1999-2000 campaign. Very useful 
conversations with Peter Gorham, Jaime Alvarez-Mu\~niz, and David 
Saltzberg are also appreciated and acknowledged. We are indebted to Dan 
DePardo and Dilip Tammana (University of Kansas Remote Sensing Lab) for 
their help in setting up and operating the KU Antenna Testing Range, and 
Alan Hase and Bruce Janus (University of Kansas Dept. of Physics) for 
their superb machine shop work. We also thank Checkers Supermarket of 
Lawrence, KS for allowing use of their vegetable freezer as an 
environmental testing chamber.

%% file: astroph_results_7_2.bbl
\begin{thebibliography}{99}
\bibitem{NEMO} NEMO: {\tt nemoweb.lns.infn.it/project.htm}.
\end{thebibliography}
